\documentstyle[12pt]{article}
\begin{document} 
\thispagestyle{empty} 
\begin{flushright}
UA/NPPS-10-04\\
\end{flushright}

\begin{center}
{\large{\bf
QUARK-HADRON CRITICAL POINT CONSISTENT \\
WITH BOOTSTRAP AND LATTICE QCD\\}}
\vspace{2cm} 
{\large N. G. Antoniou, F. K. Diakonos and A. S. Kapoyannis*}\\ 
\smallskip 
{\it Department of Physics, University of Athens, 15771 Athens, Greece}\\ 

\vspace{1cm}

\end{center}
\vspace{0.5cm}
\begin{abstract}
The critical sector of strong interactions at high temperatures is explored
in the frame of two complementary paradigms: Statistical Bootstrap for the
hadronic phase and Lattice QCD for the Quark-Gluon partition function.
A region of thermodynamic instability of hadronic matter was found, as a
direct prediction of Statistical Bootstrap. As a result, critical end point
solutions for nonzero chemical potential were traced in the phase diagram
of strongly interacting matter, consistently with chiral QCD and recent
lattice calculations. The relevance of these solutions for current and
future experiments with nuclei at high energies is also discussed.
\end{abstract}

\vspace{3cm}


*Corresponding author.
{\it E-mail address:} akapog@nikias.cc.uoa.gr (A. S. Kapoyannis)

\newpage
\setcounter{page}{1}

{\large \bf 1. Introduction}

Quantum Chromodynamics is unquestionably the fundamental theory of strong
interaction. However, the nonperturbative aspects of QCD belong still to a
field of intense investigation. In fact, most of the properties of the
hadronic world cannot be extracted yet from first QCD principles and the
partition function of interacting hadrons produced and thermalized in
high-energy collisions can only be determined within specific models of
varying degrees of limitation.

In this paper we employ Statistical Bootstrap, including volume corrections
for the finite size of hadrons, in order to describe the hadronic phase. The
virtue of this description is associated with the fact that, among Statistical
models of hadrons, it is the only one which predicts the onset of a phase
transition in strongly interacting matter and therefore it is
compatible with the properties of the QCD vacuum at high temperatures [1].
The aim of this work is to pursue the search for this compatibility with QCD
even further, asking whether Statistical Bootstrap of hadrons is compatible
with the existence of a critical endpoint in strongly interacting matter, at
high temperature and finite (nonzero) baryonic chemical potential. The
existence of such a critical point in the phase diagram is required by QCD
in extreme conditions [2] and it is the remnant of chiral QCD phase transition
[2].

The basic ingredients in our approach are (a) the hadronic partition function
extracted from the equations of Statistical Bootstrap and (b) the equation of
state of the quark-gluon phase given by recent QCD studies on the lattice
[3-6]. Our principal aim from the matching of these two descriptions is to trace
the formation of a critical point in the quark-hadron phase transition with a
mechanism compatible both with Statistical Bootstrap and Lattice QCD.

In Section 2 the principles and basic equations of Statistical Bootstrap,
leading to the partition function of hadronic phase, are briefly summarized
and the appearance, within this framework, of a thermodynamic instability is
discussed in detail. This instability is the origin of the formation of a
critical point in the bootstrap matter. In Section 3 the partition function
of the quark-gluon phase is extracted from recent lattice calculations of the
pressure of QCD matter [3]. In Section 4 the above two descriptions of
strongly interacting matter are exploited in a search for a critical point
in the phase diagram, the location of which is fixed by a set of equations
(21-25). Finally, in Section 5 our conclusions including the limitations
of our approach are presented whereas in the Appendix certain technical
points are presented concerning the evaluation of the quark-gluon partition
function (part A) and the full form that acquires the equation of maximum
hadronic pressure (part B).

\vspace{1cm}

{\large \bf 2. Thermodynamic Instability in Statistical Bootstrap}

The main attribute of the Statistical Bootstrap (SB) [7-10] is that it
describes thermodynamically hadronic systems which are interacting.
Generally as far as its thermodynamic behaviour a system of strongly
interacting entities may be considered as a collection of particles with
the explicit form of the complex interaction among themselves. Another way
to deal with the problem is to assume that the strongly interacting entities
form a greater entity with certain mass. Then, if this mass is known, the
greater entities may be treated as non-interacting and the simple
thermodynamic description of an ideal system of particles may be applied.
The SB adopts the second approach and assumes that the strongly interacting
hadrons form greater clusters, called ``fireballs''. The problem then is
moved to determine the mass of these clusters or to evaluate their mass
spectrum $\tilde{\tau}(m^2)dm^2$, which is equal to the number of
discrete hadronic states in the mass interval $\{m,m+dm\}$. The solution is
based on the assumption that the fireballs may consist of smaller fireballs
with the {\it same} mass spectrum or ``input'' particles which may not be
divided further. The integral bootstrap equation then reads [11-14]

\[
\tilde{B}(p^2)\tilde{\tau}(p^2,b,s,\ldots)=
\underbrace{g_{b,s,\ldots}\tilde{B}(p^2)\delta_0(p^2-m_{b,s,\ldots}^2)}_
{input\;term}
+\sum_{n=2}^\infty\frac{1}{n!}\int\delta^4\left(p-\sum_{i=1}^np_i\right)
\cdot
\]
\begin{equation} 
\cdot\sum_{\{b_i\}}\delta_K\left(b-\sum_{i=1}^n b_i\right) 
\sum_{\{s_i\}}\delta_K\left(s-\sum_{i=1}^n s_i\right) \ldots 
\prod_{i=1}^n\tilde{B}(p_i^2)\tilde{\tau}(p_i^2,b_i,s_i,\ldots)d^4p_i\;. 
\end{equation} 

Equation (1) also imposes conservation of four-momentum $p$ and
any kind of existing additive quantum number (baryon number $b$, strangeness
$s$, etc.) between a fireball and its constituents fireballs.
The masses $m_{b,s,\ldots}$ are the ``input'' particle masses which
constitute the smaller fireballs that can be formed.

The appropriate volume of a fireball is considered to be carried with it,
$V^{\mu}=\frac{V}{m}p^{\mu}$.
The imposition of the requirement that the sum of volumes of the constituent
fireballs has to be equal to the volume of the large fireball, as well as,
momentum conservation lead to the fact that all fireballs posses the same
volume to mass ratio. This ratio can be connected to the MIT bag constant,
$\frac{V}{m}=\frac{1}{4B}$.
The correct counting of states of a fireball involves,
apart from its mass spectrum which is of dynamical origin,
a kinematical term $\tilde{B}$
\begin{equation}
\tilde{B}(p^2)=\frac{2V^{\mu}p_{\mu}}{(2\pi)^3}=\frac{2Vm}{(2\pi)^3}\;.
\end{equation}

It can be noted in (1) that the mass spectrum is accompanied by the term
$\tilde{B}$. Thus the bootstrap equation remains unchanged if $\tilde{\tau}$
and $\tilde{B}$ are redefined in such way so
\begin{equation}
\tilde{B}\tilde{\tau}=\tilde{B}'\tilde{\tau}'\;,\;\;\;
\tilde{B}'=\frac{\tilde{m}^{4-\alpha}}{m^{4-\alpha}}\tilde{B},\;
\tilde{\tau}'=\frac{m^{4-\alpha}}{\tilde{m}^{4-\alpha}}\tilde{\tau}\;,
\end{equation}
where $\alpha$ is a free real number and $\tilde{m}$ a free parameter with
the same dimensions as the mass. The effect of such transformations will
become apparent when we turn to the thermodynamic description of the system
of fireballs.

The bootstrap equation can be simplified after performing a series of
Laplace transformations to acquire the form
\begin{equation} 
\varphi(\beta,\{\lambda\})=2G(\beta,\{\lambda\}) 
-\exp [G(\beta,\{\lambda\})]+1\;. 
\end{equation}
In the last equation $G(\beta,\{\lambda\})$ is the Laplace transform of the
mass spectrum with the accompanying kinematical term
\[
G(\beta,\{\lambda\})=
\sum_{b'=-\infty}^{\infty}\lambda_b^{b'}
\sum_{q'=-\infty}^{\infty}\lambda_q^{q'}
\sum_{s'=-\infty}^{\infty}\lambda_S^{s'}
\sum_{|s'|=0}^{\infty}\gamma_s^{|s'|}
\int e^{-\beta^\mu p_\mu}\tilde{B}'(p^2)\tilde{\tau}'(p^2,b',q',s',|s'|)dp^4\;,
\]
\begin{equation} 
\hspace{1cm}=\frac{2\pi}{\beta}\int_0^{\infty} m\tilde{B}'(m^2) \tilde{\tau}'
(m^2,\{\lambda\})K_1(\beta m)dm^2 
\end{equation} 
and $\varphi(\beta,\{\lambda\})$ the Laplace transform of the input
term
\begin{equation}
\varphi(\beta,\{\lambda\})=
\sum_{b'=-\infty}^{\infty}\lambda_b^{b'}
\sum_{q'=-\infty}^{\infty}\lambda_q^{q'}
\sum_{s'=-\infty}^{\infty}\lambda_S^{s'}
\sum_{|s'|=0}^{\infty}\gamma_s^{|s'|}
\int e^{-\beta^\mu p_\mu}g_{b'q's'|s'|}\tilde{B}'(p^2)\delta_0(p^2-m_{b'q's'|s'|}^2)
dp^4\;. 
\end{equation} 
The masses $m_{bqs|s|}$ correspond to the masses of the input
particles, which in this paper will be all the known
hadrons with masses up to 2400 MeVs, the $g_{bqs|s|}$ are degeneracy factors
due to spin and the $\lambda$'s are the fugacities of the input particles.
Here we have used the extended version of SB [14], where the states are
characterised by the set of fugacities relevant to baryon number, $\lambda_b$,
electric charge, $\lambda_q$, strangeness, $\lambda_s$ and partial
strangeness equilibrium, $\gamma_s$ (in the following we shall refer to this
set of fugacities as $\{\lambda\}$, for short).

The bootstrap equation defines the boundaries of the hadronic phase
since it exhibits a singularity at the point
\begin{equation}
\varphi(\beta,\{\lambda\}) = \ln4-1\;.
\end{equation} 

From the physical point of view the singularity is connected with the behaviour
of the mass spectrum as the mass tends to infinity
\begin{equation} 
\tilde{\tau}(m^2,\{\lambda\})\stackrel{m\rightarrow\infty} 
{\longrightarrow} 
2C(\{\lambda\})m^{-\alpha-1} \exp [m\beta^*(\{\lambda\})]\;, 
\end{equation} 
where $\beta=T^{-1}$ and $\beta^*$ corresponds to the maximum inverse
temperature. After a certain point, as temperature rises, it is more preferable
for the system to use the given energy in producing more hadronic states
(since their number rises exponentially) than in increasing the kinetic energy
of the already existing states.

In order to turn to the thermodynamics it is necessary to consider the
fireball states in an external volume $V^{ext}$. This volume must 
be distinguished from the physical volume which is carried by each
fireball.
The partition function of the pointlike interacting hadrons is then
\[ 
\ln Z(V^{ext},\beta,\{\lambda\})=\hspace{12cm} 
\]
\begin{equation}
\sum_{b'=-\infty}^{\infty}\lambda_b^{b'}
\sum_{q'=-\infty}^{\infty}\lambda_q^{q'}
\sum_{s'=-\infty}^{\infty}\lambda_S^{s'}
\sum_{|s'|=0}^{\infty}\gamma_s^{|s'|}
\int \frac{2V^{ext}_{\mu}p^{\mu}}{{(2\pi)}^3}
\tilde{\tau}'(p^2,b',q',s',|s'|) e^{-\beta^\mu p_\mu} dp^4\;.
\end{equation}
Every choice for the mass spectrum (or for the exponent
$\alpha$) in (3) leads to different partition function and so to different
physical behaviour of the system.
The usual SB choice was $\alpha=2$, but more advantageous is the choice
$\alpha=4$. With this choice a better physical behaviour is achieved as the 
system approaches the hadronic boundaries. Quantities like 
pressure, baryon density and energy density, even for 
point-like particles, no longer tend to infinity, as the system 
tends to the bootstrap singularity.
It also allows for the bootstrap singularity to be reached in the 
thermodynamic limit [15], a necessity imposed by the Lee-Yang theory.
Another point in favour of the choice $\alpha=4$ comes from the extension of
SB to include strangeness [11,12]. 
The strange chemical potential equals zero in the quark-gluon phase. With
this particular choice of $\alpha$, $\mu_s$ acquires smaller positive values
as the hadronic boundaries are approached.
With the choice $\alpha=4$ the partition function can be written down and
for point-like particles it assumes the form
\begin{equation}
\ln Z(V^{ext},\beta,\{\lambda\})=
\frac{4BV}{\beta^3}\int_{\beta}^{\infty} x^3 G(x,\{\lambda\})dx \equiv
Vf(\beta,\{\lambda\})\;.
\end{equation} 
For $\alpha=4$ the input term acquires the form
\begin{equation} 
\varphi(\beta,\{\lambda\})=\frac{1}{2\pi^2 \beta B} 
\sum_{\rm a} (\lambda_{\rm a}(\{\lambda\})+\lambda_{\rm a}(\{\lambda\})^{-1})
\sum_i g_{{\rm a}i} m_{{\rm a}i}^3 K_1 (\beta m_{{\rm a}i})\;\;, 
\end{equation} 
where ``${\rm a}$'' represents a particular hadronic family, ``$i$'' the
hadrons in the family with different masses and $B$ is the energy density of
the vacuum (MIT bag constant).

By including corrections due to the finite size of hadrons (Van der
Waals volume corrections) the repulsive part of the interaction is taken into
account.
The negative contributions to the volume can be
avoided if the following grand canonical pressure partition function is used
\begin{equation} 
\pi(\xi,\beta,\{\lambda\})=\frac{1}{\xi-f(\beta+\xi/4B,\{\lambda\})}\;,
\end{equation} 
which is the Laplace transformed partition function with respect to
volume [16] ($\xi$ is the Laplace conjugate variable of the volume).
All values of $\xi$ are allowed if Gaussian regularization 
is performed [16].
In the following we shall consider systems without any external forces
acting on them and so we shall fix $\xi=0$ [15,16].
The density and the pressure $P$ of the thermodynamic system can
be obtained through the pressure grand canonical partition function (12)
for $\xi=0$
\begin{equation}
\nu_{HG}(\xi=0,\beta,\{\lambda\})=\lambda
\frac{\partial f(\beta,\{\lambda\})}{\partial \lambda}
\left[1-\frac{1}{4B}
\frac{\partial f(\beta,\{\lambda\})}{\partial \beta}\right]^{-1}\;,
\end{equation} 
where $\lambda$ is the fugacity corresponding to the particular density, and
\begin{equation}
P_{HG}(\xi=0,\beta,\{\lambda\})=\frac{1}{\beta}
f(\beta,\{\lambda\})
\left[1-\frac{1}{4B}\frac{\partial f(\beta,\{\lambda\})}{\partial \beta}
\right]^{-1}\;.
\end{equation} 

Though volume is no longer an active variable of the system it can be
calculated for given baryon density and $\nu_B$ (evaluated through (13)) and
baryon number $<B>$ which is a conserved quantity. The volume
would be retrieved through the relation
\begin{equation}
<V>=\frac{<B>}{\nu_B}\;.
\end{equation} 

With the use of SB in order to describe interacting hadronic systems we
can trace the possibility of a phase transition. 
The study of the pressure-volume isotherm curve is then necessary. 
When this curve is calculated a region of instability is revealed.
In fact, this curve has a part (near the boundaries of the hadronic domain) where
pressure decreases while volume decreases also (see Fig. 1).
Such a behaviour is a signal of a {\it first order phase transition} which in
turn can be mended with the use of a {\it Maxwell construction}.

This behaviour is due to the formation of bigger and bigger clusters as the
system tends to its boundaries in the phase diagram. In that way the effective number of particles
is reduced, resulting, thus, to a decrease of pressure.
To show that this instability in the $P-V$ curve is the result of the
attractive part of the interaction included in the SB we shall calculate a
similar curve using the Ideal Hadron Gas (IHG) model with Van der Waals volume
corrections (repulsive part of interaction). The logarithm of the partition
function of IHG (corresponding to (10)) is
\begin{eqnarray} 
\ln Z_{p\;IHG}(V,\beta,\{\lambda\})&&\hspace{-0.1cm} \equiv
Vf_{p\;IHG}(\beta,\{\lambda\})= \nonumber \\
\frac{V}{2\pi^2\beta}
&&\hspace{-0.1cm}\sum_{\rm a} [\lambda_{\rm a}(\{\lambda\})+\lambda_{\rm a}(\{\lambda\})^{-1}]
\sum_i g_{{\rm a}i} m_{{\rm a}i} K_2 (\beta m_{{\rm a}i})\;, 
\end{eqnarray}
where $g_{{\rm a}i}$ are degeneracy factors due to spin and isospin and
the index ${\rm a}$ runs to all hadronic families.
This function can be used in eq. (12) to calculate the Ideal Hadron Gas (IHG)
pressure partition function in order to include Van der Waals volume
corrections. The result is that the pressure is always found to increase as
volume decreases, for constant temperature, exhibiting no region of
instability and so no possibility of a phase transition.

The comparison of SB with the IHG (with volume corrections) is displayed in
Fig. 1, where $\nu_0$ is the normal nuclear density $\nu_0=0.14\;fm^{-3}$.
In both cases (SB or IHG) the constraints $<S>=0$ (zero strangeness)
and $<B>=2<Q>$ (isospin symmetric system, i.e. the net number of $u$ and $d$
quarks are equal) have been imposed. Also strangeness is fully equilibrated
which accounts to setting $\gamma_s=1$.

\vspace{1cm}    

{\large \bf 3. The partition function of Quark Matter}

Having a description of the hadronic phase at hand, it is necessary to
proceed with the thermodynamical behaviour of the quark-gluon phase. 
Lattice calculations of the pressure of the quark-gluon
state have been performed at finite chemical potential in
[3,5]. These publications include calculations for 
rather heavy $u$, $d$ quark masses.
The mass of the $u$, $d$ quarks is about 65 MeV and the strange $s$
quark 135 MeV [3,5]. The calculated pressure of the quark-gluon phase ($P/T^4$)
at $\mu_B=0$
is plotted against the ratio of temperature to the transition temperature
of quark matter at zero baryon chemical potential $T/T_c$ in Fig. 2 of [3]. The
temperature $T_c$ will be denoted as $T_{0\;QGP}$ in the following.
The results of this graph are extrapolated to the continuum limit by
multiplying the raw lattice results with a factor $c_p=0.518$ [3].

The lattice calculations for finite chemical potential can be
summarised in Fig. 3 of [3], where the difference of pressure at non-zero
chemical potential and the pressure at zero chemical potential
($\Delta p/T^4 = [P(\mu \neq 0,T)-P(\mu = 0,T)]/T^4$) is plotted against
$T/T_c$. Again the results of this graph are extrapolated to the continuum
limit by multiplying the raw lattice results with a factor $c_{\mu}=0.446$ [3].
In this graph there are plotted five curves which correspond to baryon
chemical potential of 100, 210, 330, 410, 530 MeV.

With the use of Figs. 2, 3 in [3], it is possible to calculate in principle
the pressure of the quark-gluon phase at any temperature and baryon chemical
potential. The pressure is important, because knowledge of the
pressure is equivalent to the knowledge of the partition 
function of the system in the grand canonical ensemble
\begin{equation}
\ln Z_{QGP}(V,T,\mu_B)=\frac{V}{T} P(T,\mu_B)
\end{equation}

In order to have a complete description of the dependence of the pressure on
the temperature and the chemical potential we use two sets of fitting
functions. For constant chemical potential the pressure as a function of
$T/T_c$ is fitted through
\begin{equation}
f(x)=\frac{a_1}{x^{c_1} \left[ \exp\left(\frac{b_1}{x^{d_1}}\right)-1\right]^{f_1}}+
\frac{a_2}{x^{c_2} \left[ \exp\left(\frac{b_2}{x^{d_2}}\right)-1\right]^{f_2}},
\end{equation}
where $a_i,b_i,c_i,d_i,f_i\;(i=1,2)$ depend on $\mu_B$, while for constant
temperature the corresponding fit of the pressure as a function of $\mu_B$ is
given by
\begin{equation}
g(x)=a+b \exp(c x^d),
\end{equation}
where $a,b,c,d$ depend on the temperature ratio $T/T_c$. The fitting
procedure has to be performed in a self-consistent way and subsequently it is
straightforward to
evaluate the partition function as well as its derivatives with respect to
$\mu_B$ and $T$ at any given point $(T_1,\mu_{B\;1})$. In particular to
evaluate physical observables connected with the partition
function and drive numerical routines the partial derivatives of the
pressure up to second order with respect to temperature and fugacity have to
be evaluated. These derivatives are then given in part A of the Appendix.

In Fig. 2 we have reproduced the quark-gluon pressure as a function of the
temperature for constant baryon chemical potential. The squares are points
directly measured from the graphs of Fodor {\it et. al} and the lines
represent the calculation with the fits which has been performed on these
points, via eq. (18).
Fig. 3 is a graph similar with Fig. 2, but we have focused on the area which
is useful for our calculations, the area where the matching with the hadronic
phase will be performed.
Fig. 4 is a reproduction of the Fodor {\it et. al} quark-gluon pressure as
a function of the baryon chemical potential for constant temperature. The
necessary fits have been performed with the use of eq. (19).

\vspace{1cm}

{\large \bf 4. The critical point in the phase diagram}

After developing the necessary tools to handle the thermodynamic description
for the quark-gluon phase as it is produced from the lattice, we can search
for the possibility of a quark-hadron phase transition. We shall choose
$\xi=0$ [15] and so the only parameter left, in relevance with the hadronic
side, would be the maximum temperature at zero baryon chemical potential
$T_{0\;HG}$. We can, also, allow the critical temperature of quark-gluon
state at zero baryon chemical potential $T_{0\;QGP}$ as free parameter.
The fact that $T_{0\;HG}$ and $T_{0\;QGP}$ are treated as two separate
parameters does not include a contradiction, since $T_{0\;HG}$ is the
maximum temperature that the hadronic phase can exist, in the bootstrap
formalism, at $\mu_B=0$. This does not exclude the fact that the phase
transition to the quark-gluon phase can take place at a smaller temperature
at zero chemical potential. So it can happen that in general
$T_{0\;QGP}<T_{0\;HG}$ and this is the case as it will become evident in
the following. Also, the region of small baryon chemical potentials belongs,
according to the predictions of QCD, to the cross over region, where no
distinct separation between the hadronic and the quark state exists.

Then, if values for these parameters are chosen, it is possible to calculate
for a specific temperature the pressure isotherms of Hadron Gas and QGP.
Assuming that the baryon number is a conserved quantity to both phases, the
equality of volumes is equivalent to the equality of baryon densities.
and so the connection of the isotherms of the two phases is possible through 
the relation
\begin{equation}
<V_{HG}>=<V_{QGP}>\Leftrightarrow
\frac{<B_{HG}>}{\nu_{B\;HG}}=\frac{<B_{QGP}>}{\nu_{B\;QGP}}\Leftrightarrow
\nu_{B\;HG}=\nu_{B\;QGP}
\end{equation}
The graph of the pressure-volume isotherm can be drawn if the pressure is
plotted against the inverse baryon density.
Then at the point where the isotherms of the two phases meet we have
equal volumes for equal pressures.

Tracing the point where the isotherms of two phases meet, what is found 
is that at a low temperature the QGP and SB Pressure-Volume isotherms
intersect at a point where the Hadron Gas pressure is decreasing while volume 
decreases. The resulting pressure-volume curve includes an unstable part
which has to be repaired through a suitable Maxwell construction. This
curve includes a region where a first-order transition takes place.
As temperature increases, a certain temperature is found for which the QGP
and SB isotherms meet at a point which is, also, the maximum Hadron Gas
pressure for this temperature. In that case no Maxwell construction is 
needed and since this point is located at finite volume or not zero baryon
density (equivalently not zero chemical potential) it can be associated with
the QCD critical point.
As temperature rises more, the QGP and SB isotherms meet at a point which
corresponds to even greater volume. Again no Maxwell construction is 
needed and since the resulting pressure-volume isotherm always increases
while volume decreases the meeting point of the two phases belongs to the
crossover region.

A graph that summarises the situations met in the pressure volume isotherms
of hadronic and quark systems is Fig. 5. In this figure the hadronic isotherms
have been calculated for $T_{0\;HG} \simeq 177$ MeV (MIT bag constant
$B^{1/4} \simeq 222$ MeV) while the quark-gluon isotherms for
$T_{0\;QGP} \simeq 152$ MeV. For $T=161.3$ MeV a Maxwell construction is
needed to remove the instability from the resulting curve. The horizontal
line defines a partition with two equal surfaces (shaded) and represents the
final pressure-volume curve after the completion of the Maxwell construction. 
At the temperature $T=162.1$ MeV the two curves meet at the point of
maximum hadronic pressure and so a critical point is formed at finite volume.
At a greater temperature ($T \simeq 162.3$ MeV) the quark pressure-volume
curve meets the hadronic one at a point away from the area of instability
(crossover area).

To locate the critical point numerically with the use of the lattice
partition function, for given parameters $T_{0\;HG}$ and $T_{0\;QGP}$,
the conditions have to be determined for which the SB pressure is 
equal to the QGP pressure at the same volume,  
corresponding to the maximum SB pressure. Setting the factor of
partial strangeness equilibrium $\gamma_s=1$ a hadronic state is
characterised by the set of thermodynamic variables
$(T, \lambda_u, \lambda_d, \lambda_s)$, while a quark-gluon state
evaluated by the lattice [3] is characterised by the two variables
$(T, \lambda_q')$. The $u$ and $d$ quarks are characterised by the same
fugacity $\lambda_u'=\lambda_d'=\lambda_q'=\lambda_B'^{1/3}$. To
evaluate the unknown variables we have to solve the following set of
non-linear equations.
\begin{equation}
\nu_{B\;SB}(T,\lambda_u,\lambda_d,\lambda_s)=
\nu_{B\;QGP}(T,\lambda_q')
\end{equation}
\begin{equation}
P_{SB}(T,\lambda_u,\lambda_d,\lambda_s)=
P_{QGP}(T,\lambda_q')
\end{equation}
\begin{equation}
P_{SB}(T,\lambda_u,\lambda_d,\lambda_s)=
P_{SB}(T,\lambda_u,\lambda_d,\lambda_s)_{max}
\end{equation}
\begin{equation}
\left<\;S\;\right>_{SB}(T,\lambda_u,\lambda_d,\lambda_s)=0
\end{equation}
\begin{equation}
\left<\;B\;\right>_{SB}(T,\lambda_u,\lambda_d,\lambda_s)-
2\;\left<\;Q\;\right>_{SB}(T,\lambda_u,\lambda_d,\lambda_s)=0
\end{equation}
Eq. (21) accounts for the equality of the baryon densities of the two phases
which is equivalent to the equality of volumes, since the baryon number is a
conserved quantity.
Eq. (22) is the equality of pressures of Hadron Gas and QGP.
Eqs. (21), (22) determine the point where the two pressure curves meet.
Eq. (23) requires that the meeting point of the two phases for a
certain temperature is equal to the point which maximizes the Hadron Gas
pressure and so this meeting point is the critical point. The form of this
equation is discussed in detail in the Appendix.
Eq. (24) imposes strangeness neutrality in the hadronic phase.
Eq. (25) imposes isospin symmetry to the hadronic system. The solution for
systems without such an attribute is available for the hadronic side but it
is of no use in the current study since the lattice results with only one
fugacity for the quark system have included the assumption of isospin
symmetry.

Before presenting results on the existence and location of the critical point,
certain limitations on the available parameters
$T_{0\;HG}$ and $T_{0\;QGP}$ are in order. To have a critical point which
could be associated with QCD it is necessary for this
point to be located at non zero chemical potential. So to impose limits on
the values of these parameters we have to evaluate for which choices the
critical point corresponds to zero chemical potential. At $\mu_B=0$ the
constraints $\left<S\right>_{SB}=0$ and
$\left<B\right>_{SB}-2\left<Q\right>_{SB}=0$ may be solved analytically to
give $\lambda_u=\lambda_d=\lambda_s=1$. Then, $\nu_{B\;SB}=0$ and
consequently $\nu_{B\;QGP}$ has to be set to zero, leading to  $\lambda_q'=1$.
Only, the temperature $T$, is left undetermined from the set of thermodynamic
variables of the system. For given parameter $T_{0\;HG}$ one can, then,
solve the set of eqs. (22) and (23) to determine the value of the set
$(T;T_{0\;QGP})$. The results of this calculation are depicted on Fig. 6,
where the solid line represents the connection of the two parameters for a
critical point at $\mu_B=0$. It is easy then to check that the region on
the upper left of this curve represents choices of the parameters that drive the
critical point on positive baryon chemical potential, whereas the region on
the bottom and right represents choices of the parameters for which the set of
eqs. (21)-(25) has no solution. It is evident from the slashed line, 
that it is not possible to have solution for
the critical point for the choice $T_{0\;HG}=T_{0\;QGP}$. The difference,
though, $DT=T_{0\;HG}-T_{0\;QGP}$ between the two parameters has a minimum,
which is about $5.1$ MeV and it occurs for the choice $T_{0\;HG}=140.6$ MeV
($B^{1/4}=136$ MeV) and $T_{0\;QGP}=135.5$ MeV.

The solutions for the position of the critical point in the $(T,\mu_B)$ plane
for specific choice of the parameters are presented in Fig. 7. The solid lines
represent solutions with fixed $T_{0\;QGP}$ and varying $T_{0\;HG}$, whereas
the slashed lines represent the opposite situation. It is evident that an
increase of $T_{0\;HG}$ at constant $T_{0\;QGP}$ leads the critical point to
higher temperature and higher baryon chemical potential. An increase of
$T_{0\;QGP}$ at constant $T_{0\;HG}$ drives the critical point at higher 
temperatures but at smaller baryon chemical potential.

Recent lattice QCD studies offer, apart from the quark-gluon pressure which
has been a basic ingredient in our approach, important results on the
existence and location of the critical point itself.
In [17] the critical point is found to reside at $T_{cr.p.}=160\pm3.5$ MeV and
$\mu_B=725\pm35$ MeV, with $T_{0\;QGP}=172\pm3$ MeV. These
calculations have the drawback that have been performed with rather
unphysical mass for $u$ and $d$ quarks which has a value about four times the
physical value.
Improved calculations have been performed in [18], where the light quark
masses have decreased by a factor of 3 down their physical values. The
critical point is found now (with $T_{0\;QGP}=164\pm3$ MeV) to be at
$T_{cr.p.}=162\pm2$ MeV and $\mu_B=360\pm40$ MeV, a value which is
considerably reduced with respect to the previous one. This point is depicted
on Fig. 7 with the full circle. In our study it can be approached with the
choice of the parameters $T_{0\;HG}=152.5$ MeV and $T_{0\;QGP}=177.5$ MeV.

In Fig. 8 we present our solution for the QGP-hadron transition line in the
$(T,\mu_B)$ plane. The parameters are chosen so as to make the critical point
to be in the same location as the lattice solution. The circle
represents our solution for the critical point and the star the lattice 
result [18]. The solid thick line represents the boundaries of the hadronic
phase (as set from the bootstrap singularity) which is close to the first
order critical line. 
The solid thin line is the first order transition line evaluated in [17] with
large $u$,$d$ quark masses. The dotted line represents the crossover in this
calculation. The corresponding solution for the critical point is also
depicted, at a large value of baryon chemical potential.
For completeness, early solutions for the critical point [2] based on different
approximate theories (Nambu-Jona-Lasinio (NJL) model or a random matrix (RM)
approach) are also shown in the same figure (NJL-RM).
We have, also depicted on the same graph the freeze-out points of different
experiments. It is evident that the recent lattice calculations [18]
set the critical point to a location easily accessible by experiments.

\vspace{1cm}

{\large \bf 5. Conclusions}

Statistical Bootstrap presents a more accurate description of the
hadronic phase than the ideal Hadron Gas, since it includes in a self
consistent way the interaction among hadrons. This interaction is crucial to
investigate critical phenomena in connection with the state of quark-gluon
plasma. Among the predictions of the bootstrap model is the limitation of the
hadronic phase and the forming of an instability in the pressure-volume
isotherm near the hadronic boundaries. This instability can be connected
with a first order quark-hadron phase transition and the existence of a
critical end point in the strongly interacting matter.

The accurate partition function of the quark-gluon phase is available from
lattice calculations, which include, though, large values of the light
quark masses. From these results the lattice partition function of the
quark state, as well as, all the necessary derivatives can be calculated,
allowing the evaluation of any physical observable.

The joining of the SB and the lattice partition function for the haronic and
the quark state respectively, allows for the determination of a critical
point at finite baryon chemical potential which can be related to the
critical point of QCD.

More recent lattice calculations [18] drive the position of the critical point
to smaller values of baryon chemical potential as the values of $u-d$ quark
masses approach their physical values. It is interesting that the current
location is situated in the $(T,\mu_B)$ plane in a region easily accessible
by the freeze-out conditions of experiments at the CERN/SPS.

With suitable choice of the temperature at zero density ($T_0$) for both the
hadronic and the quark-gluon system the position of the critical point,
determined in our study through the statistical bootstrap and the lattice
pressure, can coincide with the one directly found from lattice calculations.

In a previous work [19,20] a similar solution was found with the use of a
simplified partition function for the quark-gluon system, based on
the MIT bag model. Therefore, the basic mechanism in our approach for the
formation of a critical point in the strongly interacting matter is not
associated with the details of the quark-gluon partition function but mainly
with the instability of hadronic matter revealed by Statistical Bootstrap.
In particular the local maximum of pressure in the $P-V$ diagram (fig. 5) of
hot hadronic matter lies in the origin of the formation mechanism of the
critical end point.

In the absence of a final theory of strongly interacting matter, providing a
unified partition function for both phases, our approach is necessarily an
approximate treatment with drawbacks and limitations. In fact the continuity
of the derivatives of $P-V$ isotherms in the crossover area ($T>T_c$) is not
guaranteed (fig. 5) and the power law at the critical temperature ($T=T_c$),
dictated by the isotherm critical exponent $\delta$, is not revealed in this
treatment. Despite these limitations, the overall consistency of the critical
sector of strong interactions at high temperatures, emphasized in this
approach, provides us with extra confidence for the existence and the
location of the QCD critical end point and makes its experimental discovery
even more challenging [21].

\vspace{1cm}

{\large \bf Acknowledgements}

This work was supported in part by the Research Committee of the University
of Athens.

\vspace{1cm}

{\large \bf Appendix}

{\large \bf A.} The fitting procedures on curves of constant temperature and
chemical potential allows us to evaluate the derivatives with respect to $T$
for constant $\mu_B$ and with respect to $\mu_B$ for constant $T$. Physical
observables, however, are given as derivatives with respect to temperature
for constant fugacity or with respect to $\lambda_B$ for constant $T$. The
evalution of the latter (for the pressure) is given by
\[
\hspace{3.8cm}
\left. \frac{\partial P}{\partial \lambda_B}\right|_T=
\left. \frac{\partial P}{\partial \mu_B}\right|_T \frac{T}{\lambda_B} ,
\hspace{3cm} (A.1)
\]
\[
\hspace{3.8cm}
\left. \frac{\partial P}{\partial T}\right|_{\lambda_B}=
\left. \frac{\partial P}{\partial T}\right|_{\mu_B}+
\left. \frac{\partial P}{\partial \lambda_B}\right|_{T}
\frac{\lambda_B \mu_B}{T^2}=
\left. \frac{\partial P}{\partial T}\right|_{\mu_B}+
\left. \frac{\partial P}{\partial \mu_B}\right|_{T}
\frac{\mu_B}{T},
\hspace{1cm} (A.2)
\]
\[
\hspace{3.8cm}
\left. \frac{\partial^2 P}{\partial \lambda_B^2}\right|_{T}=
- \left. \frac{\partial P}{\partial \mu_B}\right|_{T} \frac{T}{\lambda_B^2}+
\left. \frac{\partial^2 P}{\partial \mu_B^2}\right|_{T}
\frac{T^2}{\lambda_B^2},
\hspace{3cm} (A.3)
\]
\[
\hspace{3.8cm}
\left. \frac{\partial^2 P}{\partial T^2}\right|_{\lambda_B}=
\left. \frac{\partial^2 P}{\partial T^2}\right|_{\mu_B} +
\frac{\partial^2 P}{\partial \mu_B \partial T} \frac{2 \mu_B}{T} +
\left. \frac{\partial^2 P}{\partial \mu_B^2}\right|_{T} \frac{\mu_B^2}{T^2},
\hspace{3cm} (A.4)
\]
\[
\hspace{3.8cm}
\frac{\partial^2 P}{\partial \lambda_B \partial T}=
\frac{\partial^2 P}{\partial T \partial \mu_B} \frac{T}{\lambda_B}+
\left. \frac{\partial^2 P}{\partial \mu_B^2} \right|_{T}
\frac{\mu_B}{\lambda_B} +
\left. \frac{\partial P}{\partial \mu_B}\right|_{T} \frac{1}{\lambda_B}.
\hspace{3cm} (A.5)
\]
In eqs. (A.4),(A.5) where the 2nd partial derivative with respect to two
different variables of $P$ appears, the pressure is considered as a function of
these two variables.

{\large \bf B.} The requirement of eq. (23) has to be fulfilled for a certain temperature
$T=T_1$ and in the presence of two constraints
$g_1 \equiv \left<\;S\;\right>_{SB}(T_1,\lambda_u,\lambda_d,\lambda_s)=0$
and $g_2 \equiv$
$\left<\;B\;\right>_{SB}(T_1,\lambda_u,\lambda_d,\lambda_s)-
2\;\left<\;Q\;\right>_{SB}(T_1,\lambda_u,\lambda_d,\lambda_s)=0$.
The temperature $T_1$, since the maximum of pressure is found for a certain
isotherm, may not be considered in the following as an active variable.
\[
\frac{dP_{SB}(T_1,\lambda_u,\lambda_d,\lambda_s)}{d \nu_B}=0
\Rightarrow dP_{SB}=0 \Rightarrow
\frac{\partial P_{SB}}{\partial \lambda_u} d\lambda_u+
\frac{\partial P_{SB}}{\partial \lambda_d} d\lambda_d+
\frac{\partial P_{SB}}{\partial \lambda_s} d\lambda_s=0\Rightarrow
\]
\[
\hspace{3.8cm}
\frac{\partial P_{SB}}{\partial \lambda_u} +
\frac{\partial P_{SB}}{\partial \lambda_d} \frac{d\lambda_d}{d\lambda_u}+
\frac{\partial P_{SB}}{\partial \lambda_s} \frac{d\lambda_s}{d\lambda_u}=0
\hspace{3cm} (B.1)
\]

As far the constraint $g_1$ is concerned, we have
\[
g_1(T_1,\lambda_u,\lambda_d,\lambda_s)=0\Rightarrow dg_1=0\Rightarrow
\frac{\partial g_1}{\partial \lambda_u} d\lambda_u+
\frac{\partial g_1}{\partial \lambda_d} d\lambda_d+
\frac{\partial g_1}{\partial \lambda_s} d\lambda_s=0\Rightarrow
\]
\[
\hspace{3.8cm}
\frac{\partial g_1}{\partial \lambda_d} \frac{d\lambda_d}{d\lambda_u}+
\frac{\partial g_1}{\partial \lambda_s} \frac{d\lambda_s}{d\lambda_u}=
-\frac{\partial g_1}{\partial \lambda_u}
\hspace{3cm} (B.2)
\]
Similarly for the constraint $g_2$ we have
\[
\hspace{3.8cm}
\frac{\partial g_2}{\partial \lambda_d} \frac{d\lambda_d}{d\lambda_u}+
\frac{\partial g_2}{\partial \lambda_s} \frac{d\lambda_s}{d\lambda_u}=
-\frac{\partial g_2}{\partial \lambda_u}
\hspace{3cm} (B.3)
\]
Eqs. (B.2) and (B.3) may considered as a system of two equations which
can be solved to determine $d\lambda_d / d\lambda_u$ and
$d\lambda_s / d\lambda_u$
\[
\hspace{3.8cm}\frac{d\lambda_d}{d\lambda_u}=\frac{1}{D}\left(
\frac{\partial g_1}{\partial \lambda_s}
\frac{\partial g_2}{\partial \lambda_u}-
\frac{\partial g_2}{\partial \lambda_s}
\frac{\partial g_1}{\partial \lambda_u} \right), \hspace{3cm} (B.4)
\]
\[
\hspace{3.8cm}\frac{d\lambda_s}{d\lambda_u}=\frac{1}{D}\left(
\frac{\partial g_1}{\partial \lambda_u}
\frac{\partial g_2}{\partial \lambda_d}-
\frac{\partial g_2}{\partial \lambda_u}
\frac{\partial g_1}{\partial \lambda_d} \right), \hspace{3cm} (B.5)
\]
with
\[
\hspace{3.8cm}D=\frac{\partial g_1}{\partial \lambda_d}
\frac{\partial g_2}{\partial \lambda_s}-
\frac{\partial g_2}{\partial \lambda_d}
\frac{\partial g_1}{\partial \lambda_s}. \hspace{3cm} (B.6)
\]

Eqs. (B.4) and (B.5) may now be inserted to eq. (B.1) to give
\[
\frac{\partial P_{SB}}{\partial \lambda_u} +
\frac{\partial P_{SB}}{\partial \lambda_d}
\left(
\frac{\partial g_1}{\partial \lambda_s}
\frac{\partial g_2}{\partial \lambda_u}-
\frac{\partial g_2}{\partial \lambda_s}
\frac{\partial g_1}{\partial \lambda_u} \right) \frac{1}{D} +
\frac{\partial P_{SB}}{\partial \lambda_s}
\left(
\frac{\partial g_1}{\partial \lambda_u}
\frac{\partial g_2}{\partial \lambda_d}-
\frac{\partial g_2}{\partial \lambda_u}
\frac{\partial g_1}{\partial \lambda_d} \right) \frac{1}{D}
=0 \hspace{0.3cm} (B.7)
\]
Eq. (B.7) is the form of eq. (23) for the maximum hadron pressure in a
certain isotherm. The main contribution comes from the term
$\partial P_{SB} / \partial \lambda_u$ and the result for the
maximum pressure is near to the maximum evaluated with the full
equation (B.7), which explains why only the first term was used in [19,20].

\vspace{1cm}

{\large \bf Figure Captions}

\newtheorem{f}{Fig.}
\begin{f} 
\rm Isotherm pressure-volume curve for SB and IHG (both with Van der Waals
    volume corrections using the pressure ensemble).  The SB curve is
    drawn for $T_{0\;HG}=173$ MeV ($B^{1/4}=210$ MeV).
\end{f} 
\begin{f}
\rm The pressure of the quark-gluon state divided by $T^4$ versus the ratio
    $T/T_{0\;QGP}$, for constant baryon chemical potential. The lines from
    bottom to top correspond to gradually increasing values of $\mu_B$. The
    squares represent direct measurement from the figs 2, 3 in [3] which depict
    the lattice calculation and the lines our fits on these points.
\end{f}
\begin{f}
\rm Similar graph with Fig. 2. The pressure of the quark-gluon state is
    divided by $T_{0\;QGP}^4$ and the graph is focusing on the region where
    the matching with the hadronic state will take place.
\end{f}
\begin{f}
\rm The pressure of the quark-gluon state divided by $T_{0\;QGP}^4$ versus
    the baryon chemical potential, for constant values of the ratio
    $T/T_{0\;QGP}$. The squares represent direct measurement from the figs 2, 3
    in [3] which depict the lattice calculation and the lines our fits on
    these points.
\end{f}
\begin{f}
\rm Three isotherm pressure-volume curves for Hadron Gas (using SB) and QGP
    phase (using the lattice pressure of [3]).
    The low temperature isotherm
    needs Maxwell construction, the middle temperature isotherm develops a
    critical point and the high temperature isotherm corresponds to
    crossover.
\end{f} 
\begin{f}
\rm Connection between the parameters $T_{0\;HG}$ and $T_{0\;QGP}$ (solid
    line) in order for the critical point to be developed at zero baryon
    chemical potential. The region which leads to critical point
    at positive chemical potential (upper left) and the region with no
    solution (bottom right) are depicted.
\end{f}
\begin{f} 
\rm Position of the critical point at the $(T,\mu_B)$ plane for constant
    values of $T_{0\;HG}$ and varying $T_{0\;QGP}$ (solid lines) and
    constant values of $T_{0\;QGP}$ and varying $T_{0\;HG}$ (slashed
    lines). The lattice calculated critical point in [18] is displayed by
    the full circle and the cross.
\end{f}
\begin{f}
\rm Position of the critical point (open circle) (at the $(T,\mu_B)$ plane for
    the specific choice of parameters $T_{0\;HG}=177.5$ MeV and
    $T_{0\;QGP}=152.5$ MeV as well as the first order part quark-hadron
    transition (bootstrap singularity-thick line). The stars represent the
    lattice calculated critical points in [17] and [18]. The thin line is
    reproduced from [17] and represents lattice calculation for the transition
    line. There are, also, depicted freeze-out points from different
    experiments.
\end{f}


\begin{thebibliography}{99}
\bibitem{1} N. Cabbibo, G. Parisi, Phys. Lett. B 59 (1975) 67.
\bibitem{2} F. Wilczek, e-print archive: hep-ph/0003183;\\
J. Berges, K. Rajagopal, Nucl. Phys. B 538 (1999) 215;\\
M. A. Halasz, A. D. Jackson, R. E. Shrock, M. A. Stephanov,
J. J. Verbaarschot, Phys. Rev. D 58 (1998) 096007.
\bibitem{3} Z. Fodor, S. D. Katz, K. K. Szabo,
Phys. Lett. B 568 (2003) 73, e-print archive: hep-lat/0208078.
\bibitem{4} F. Csikor, G. I. Egri, Z. Fodor, S. D. Katz, K. K. Szabo, A. I. Toth,
Contributed to Workshop on Strong and Electroweak Matter (SEWM 2002),
Heidelberg, Germany, 2-5 Oct 2002, e-print archive: hep-lat/0301027.
\bibitem{5} F. Csikor, G. I. Egri, Z. Fodor, S. D. Katz, K. K. Szabo, A. I. Toth,
Nucl. Phys. Proc. Suppl. 119 (2003) 547, e-print archive: hep-lat/0209114.
\bibitem{6} F. Csikor, G. I. Egri, Z. Fodor, S. D. Katz, K. K. Szabo, A. I. Toth,
talk given at Finite Density QCD at Nara, Nara, Japan, 10-12 July 2003,
Prog. Theor. Phys. Suppl. 153 (2004) 93, e-print archive: hep-lat/0401022.
\bibitem{7} R. Hagedorn, Suppl. Nuovo Cimento III (1965) 147. 
\bibitem{8} R. Hagedorn, J. Ranft, Suppl. Nuovo Cimento VI (1968) 169;
R. Hagedorn, Suppl. Nuovo Cimento VI (1968) 311.
\bibitem{9} R. Hagedorn, Nuovo Cimento LVI A (1968) 1027.
\bibitem{10} R. Hagedorn, J. Rafelski, Phys. Lett. B 97 (1980) 136. 
\bibitem{11} A. S. Kapoyannis, C. N. Ktorides, A. D. Panagiotou, 
J. Phys. G 23 (1997) 1921.
\bibitem{12} A. S. Kapoyannis, C. N. Ktorides, A. D. Panagiotou, 
Phys. Rev. D 58 (1998) 034009.
\bibitem{13} A. S. Kapoyannis, C. N. Ktorides, A. D. Panagiotou, 
Phys. Rev. C 58 (1998) 2879. 
\bibitem{14} A. S. Kapoyannis, C. N. Ktorides, A. D. Panagiotou, 
Eur. Rhys. J. C 14 (2000) 299.
\bibitem{15} J. Letessier, A. Tounsi, Nuovo Cimento 99A (1988) 521.
\bibitem{16} R. Hagedorn, Z. Phys. C 17 (1983) 265.
\bibitem{17}
Z. Fodor, S. D. Katz, JHEP 0203 (2002) 014,
e-Print Archive: hep-lat/0106002;\\
Z. Fodor, S. D. Katz,
talk given at Finite Density QCD at Nara, Nara, Japan, 10-12 July 2003,
Prog. Theor. Phys. Suppl. 153 (2004) 86, e-print archive: hep-lat/0401023.
\bibitem{18} Z. Fodor, S. D. Katz,
JHEP 0404 (2004) 050, e-print archive: hep-lat/0402006.
\bibitem{19} N. G. Antoniou, F. K. Diakonos, A. S. Kapoyannis,
Proc. 10th International Workshop on Multiparticle Production, Crete,
Greece 8-15 June 2002. World Scientific, p.201, Edited by: N.
G. Antoniou, F. K. Diakonos and C. N. Ktorides.
\bibitem{20} N. G. Antoniou, A. S. Kapoyannis,
Phys. Lett. B 563 (2003) 165.
\bibitem{21} N. G. Antoniou, Y. F. Contoyiannis, F. K. Diakonos,
G. Mavromanolakis, e-print archive: hep-ph/0307153;\\
N. G. Antoniou, Y. F. Contoyiannis, F. K. Diakonos, A. I. Karanikas,
C. N. Ktorides, Nucl. Phys. A 693 (2001) 799.
\end{thebibliography}
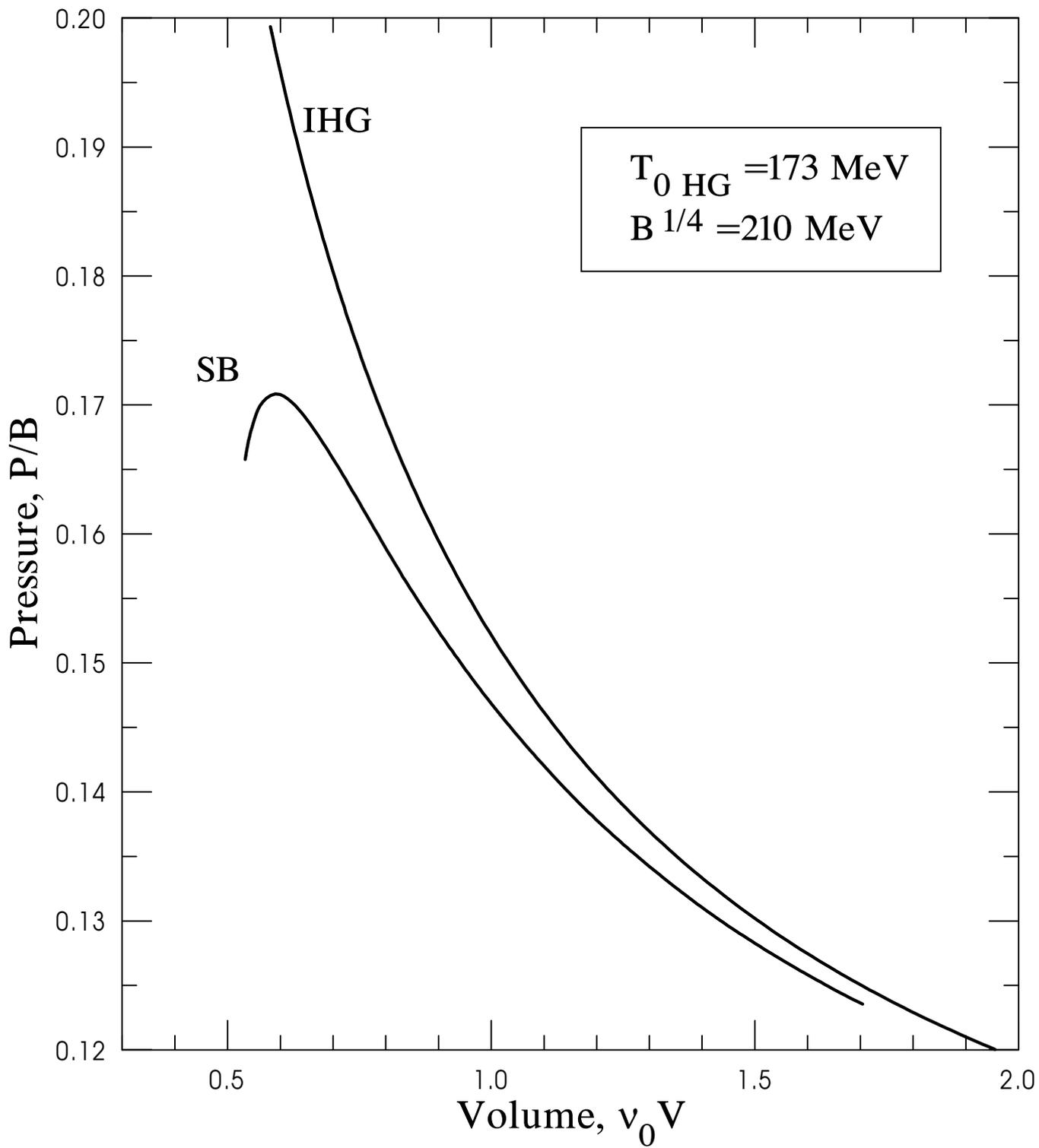
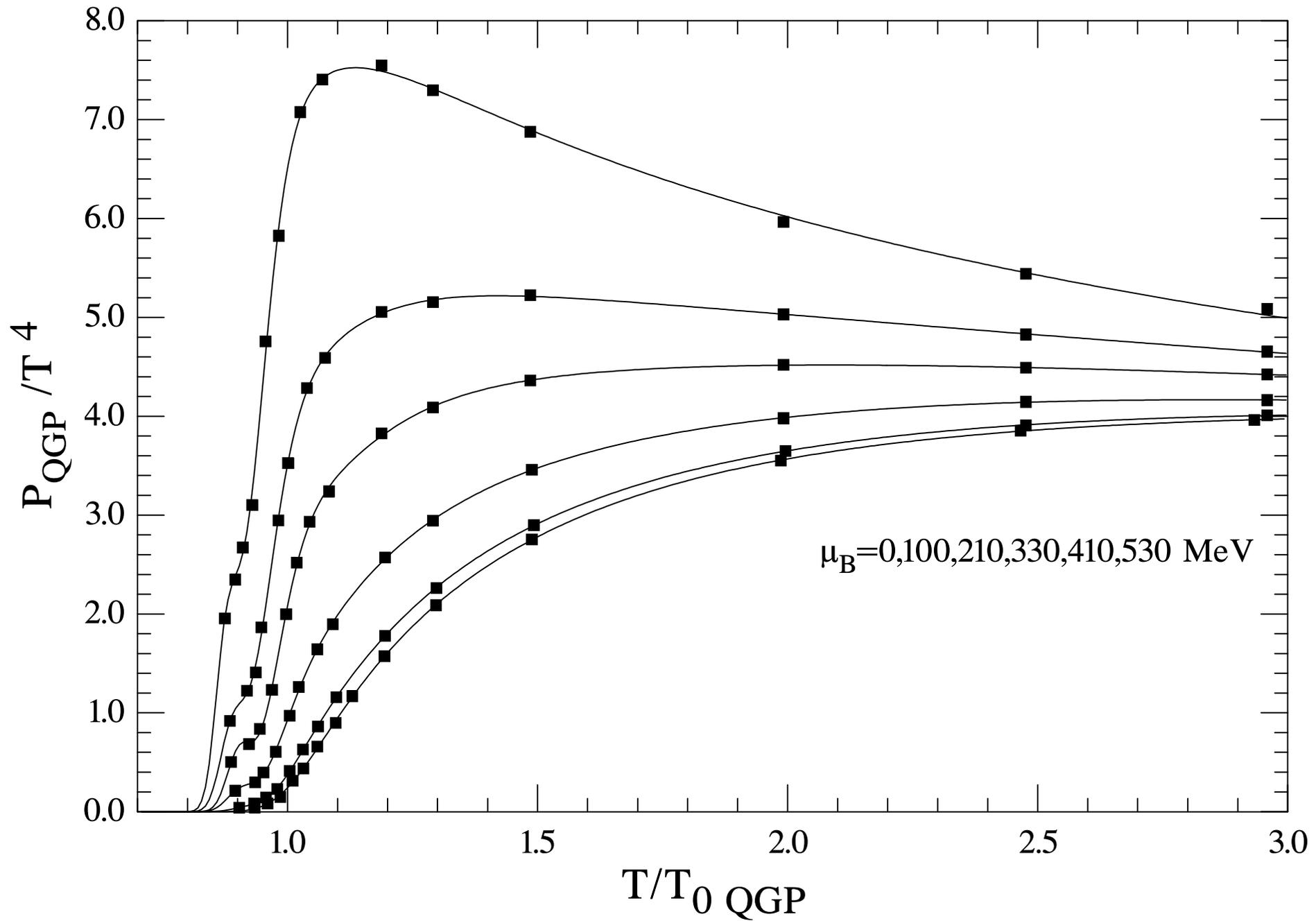
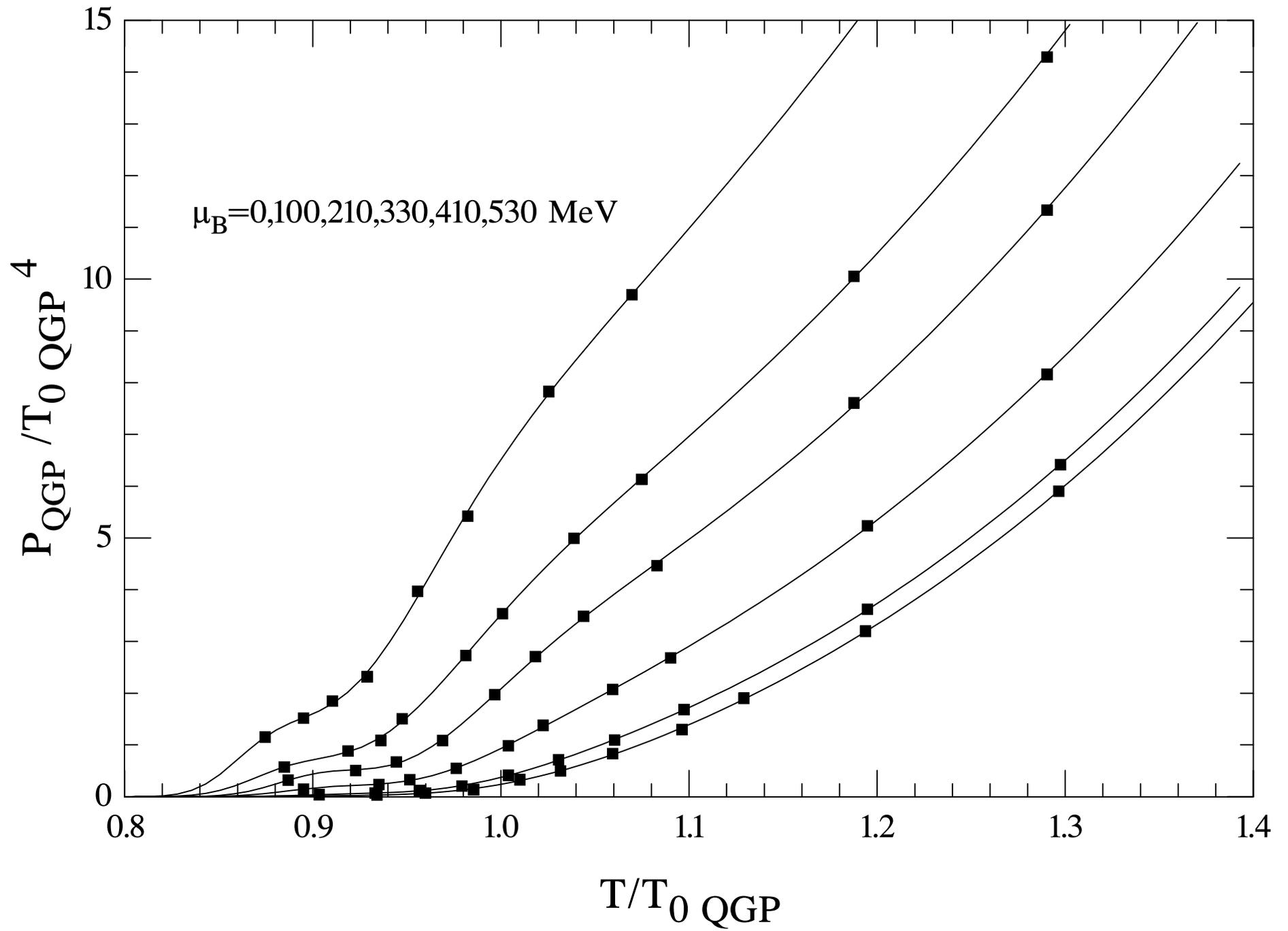
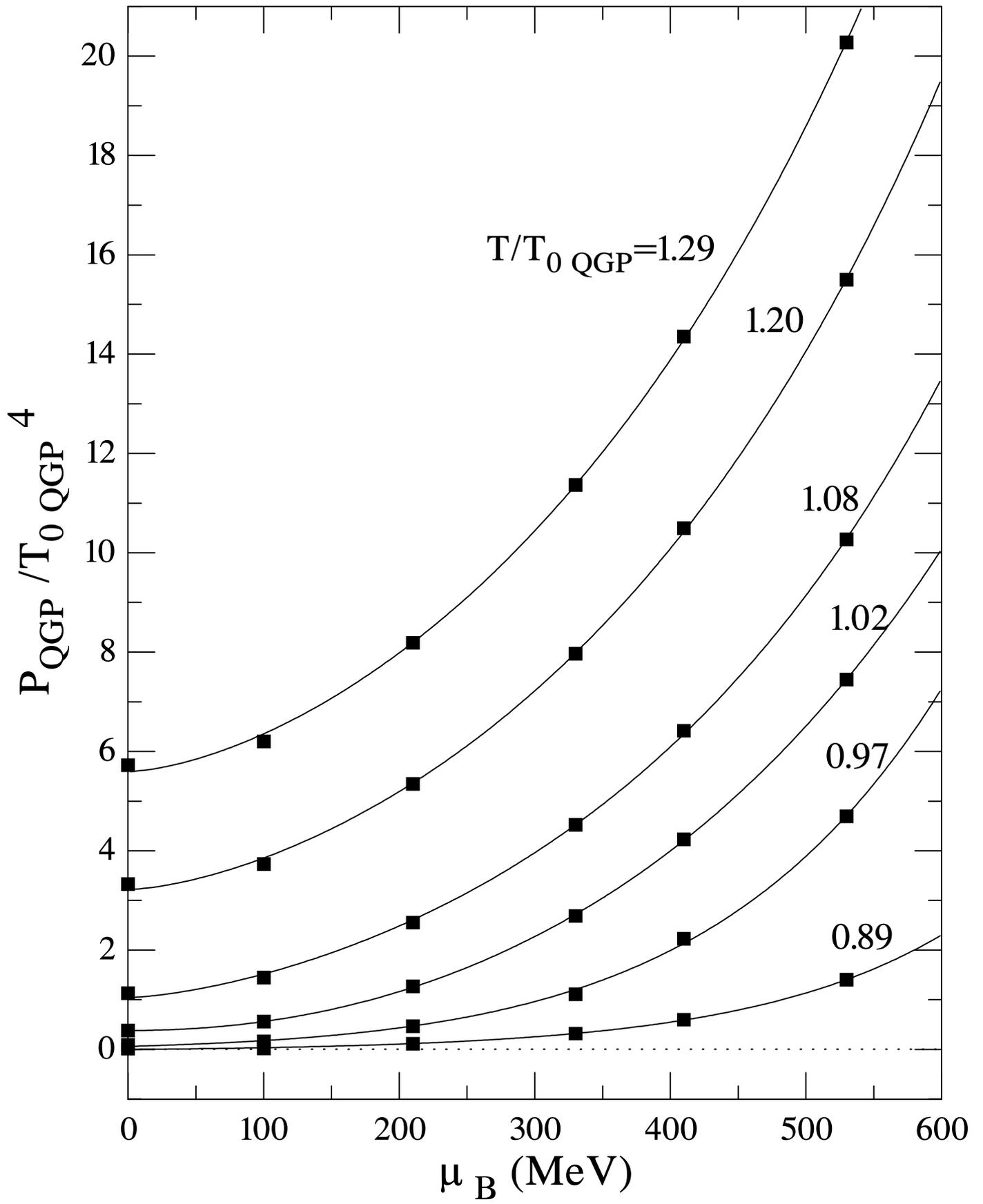
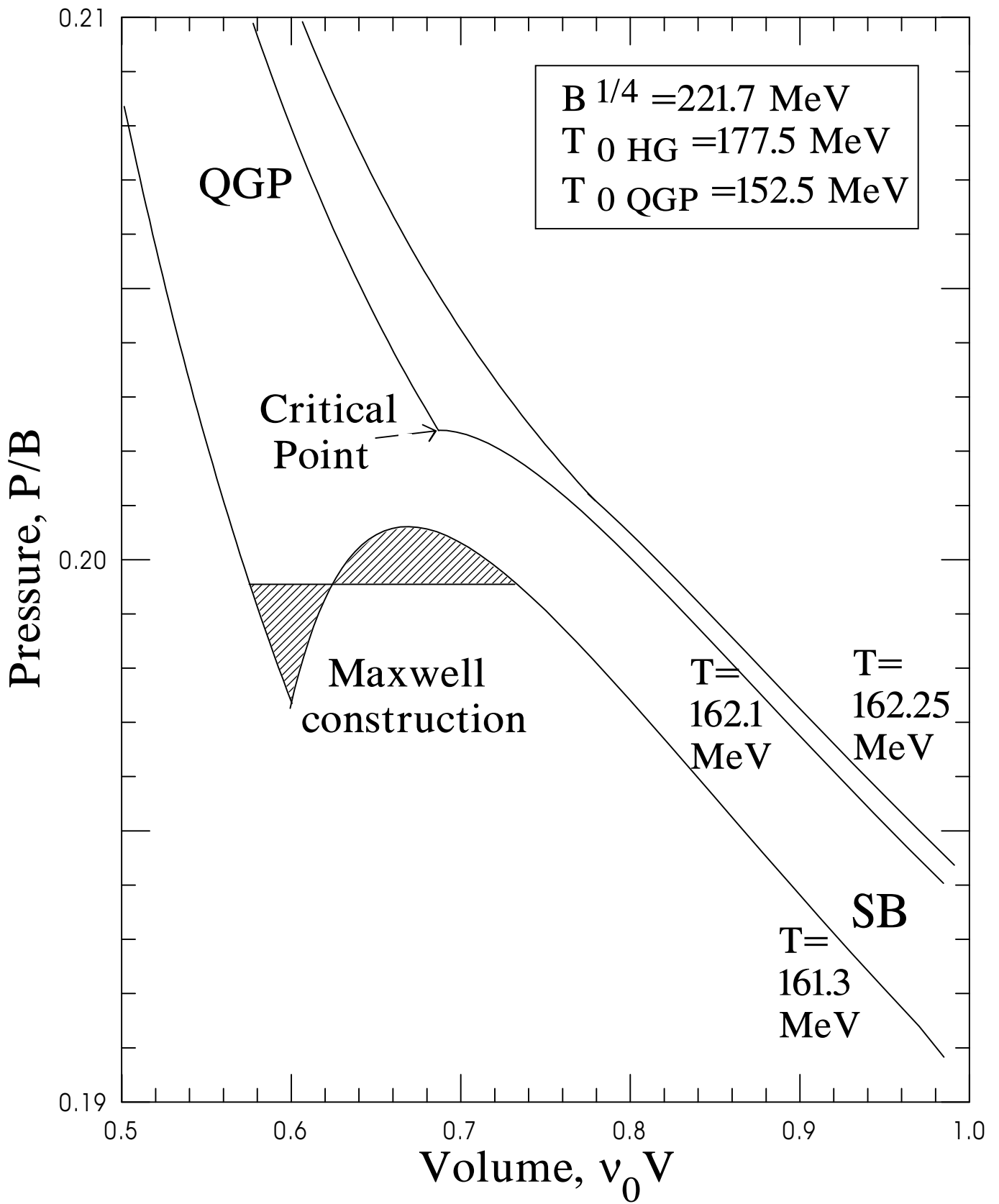
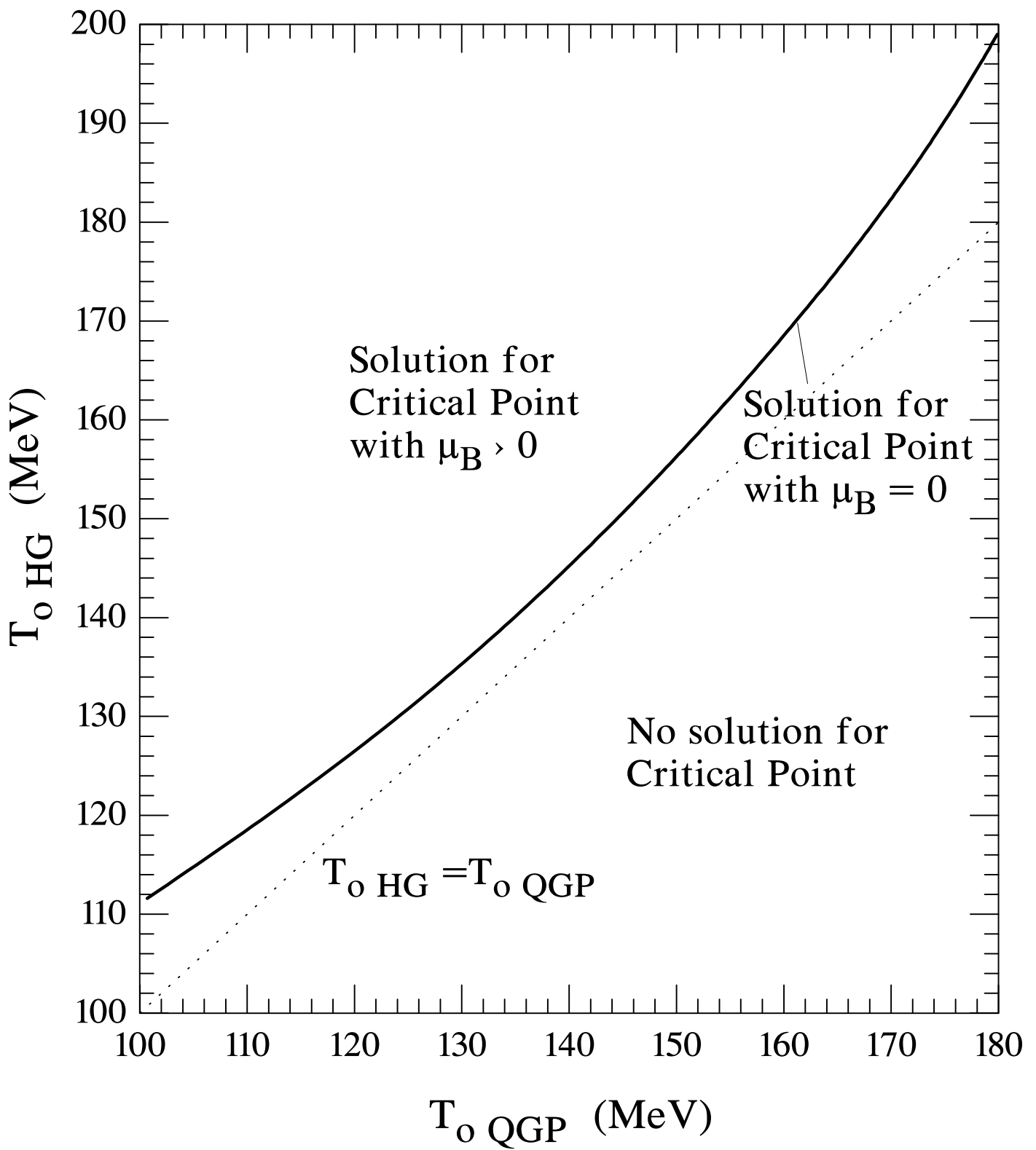
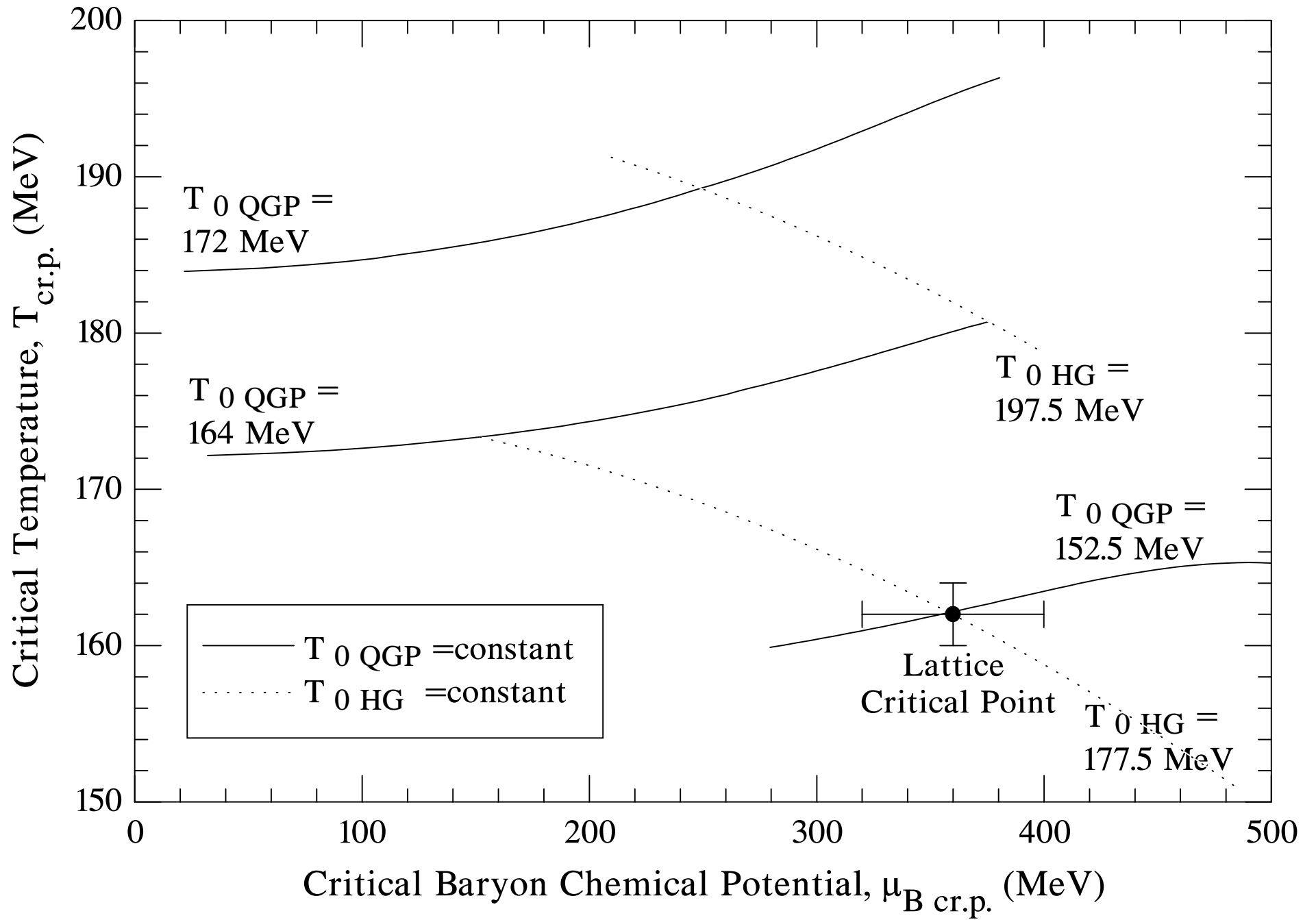
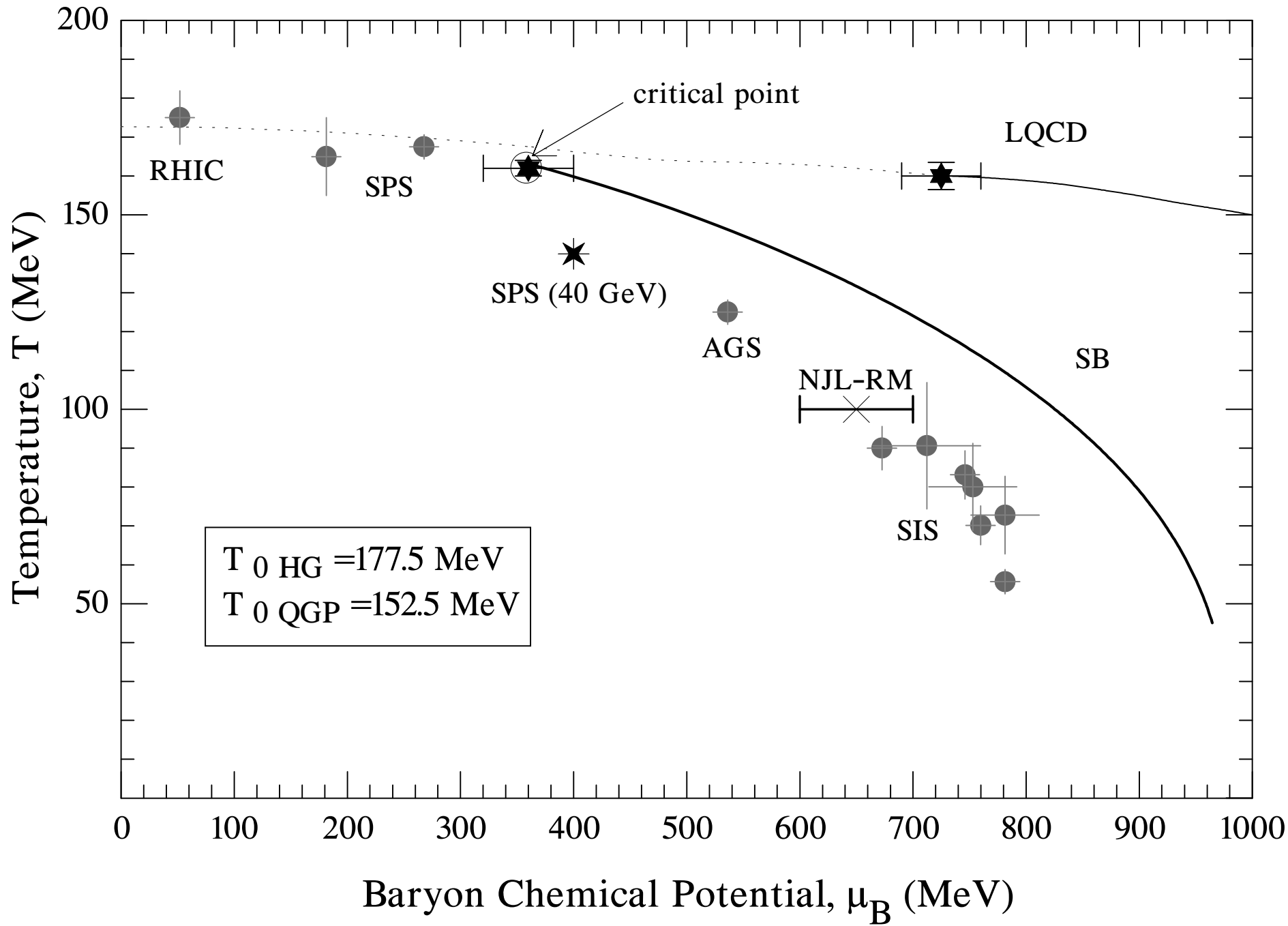
\end{document}